\documentclass[aps,twocolumn,groupedaddress,showpacs,showkeys,nofootinbib,amsfonts,eqsecnum,floatfix]{revtex4}

\usepackage{graphicx}
\usepackage{epsfig}
\usepackage{bm}

\newcommand{\tr}{{\textrm {tr}}}
\newcommand{\Tr}{{\textrm {Tr}}}

\newcommand{\D}{{\widehat D}}

\newcommand{\hb}{b^{*}}
\newcommand{\thruu}[1]{\mathrel{\mathop{#1\!\!\!\!/}}}
\newcommand{\thru}[1]{\mathrel{\mathop{#1\!\!\!/}}}

\newcommand{\I}{{\cal I}}

\newcommand{\oI}{{\overline{\cal I}}}
\newcommand{\cL}{{\cal L}}

\def\slashchar#1{\setbox0=\hbox{$#1$}
   \dimen0=\wd0 \setbox1=\hbox{/} \dimen1=\wd1
   \ifdim\dimen0>\dimen1 \rlap{\hbox to \dimen0{\hfil/\hfil}} #1
   \else  \rlap{\hbox to \dimen1{\hfil$#1$\hfil}} / \fi}

\def\D{{\bf D}}

\begin{document}

\title{Chiral Lagrangian at finite temperature from the
Polyakov-Chiral Quark Model}

\author{E. Meg\'{\i}as}
\email{emegias@ugr.es}

\author{E. \surname{Ruiz Arriola}}
\email{earriola@ugr.es}

\author{L.L. Salcedo}
\email{salcedo@ugr.es}

\affiliation{
Departamento de F\'{\i}sica At\'omica, Molecular y Nuclear,
Universidad de Granada,
E-18071 Granada, Spain
}

\date{\today}

\begin{abstract}
We analyze the consequences of the inclusion of the gluonic Polyakov
loop in chiral quark models at finite temperature. Specifically, the
low-energy effective chiral Lagrangian from two such quark models is
computed. The tree level vacuum energy density, quark condensate, pion
decay constant and Gasser-Leutwyler coefficients are found to acquire
a temperature dependence. This dependence is, however, exponentially
small for temperatures below the mass gap in the full unquenched
calculation. The introduction of the Polyakov loop and its quantum
fluctuations is essential to achieve this result and also the correct
large $N_c$ counting for the thermal corrections.  We find that new
coefficients are introduced at ${\cal O}(p^4)$ to account for the
Lorentz breaking at finite temperature. As a byproduct, we obtain the
effective Lagrangian which describes the coupling of the Polyakov loop
to the Goldstone bosons.
\end{abstract}



\maketitle

\section{INTRODUCTION}
\label{sec:intro}

At zero temperature, confinement and spontaneous breaking of chiral
symmetry emerge as distinct features of QCD. This explains the absence
of free quarks and gluons as well as the observed mass gap in the
hadron spectrum between the pseudoscalar mesons and the rest of
particles and resonances.  At a given critical temperature, lattice
simulations predict a deconfinement phase transition where chiral
symmetry is simultaneously restored (for a review see
e.g. Ref.~\cite{Karsch:1998hr} and references therein). This
remarkable coincidence between the deconfinement and chiral phase
transitions remains so far unexplained from the theoretical
side. Nevertheless, there are two extreme limits where each of the
phase transitions can be characterized by an order parameter. For
extremely light quarks, the quark condensate is used as an order
parameter for the chiral phase transition where one goes from a
non-vanishing to a vanishing value across the phase transition. In the
opposite limit of infinitely heavy quarks, the deconfinement phase
transition can be characterized by a breaking of the center symmetry
of the gauge group and the order parameter is the Polyakov loop which
evolves from a vanishing value to unity above the critical
temperature. The real situation for light quarks is in between but one
still observes sudden changes both in the chiral condensate as well as
in the Polyakov loop. In the present paper we want to address in a
quantitative manner a rather remarkable feature that arises at low
temperatures in quark models when the chiral flavour symmetry and
colour gauge center symmetry are jointly considered.

Besides the existence of a mass gap, a further outstanding consequence
of spontaneous chiral symmetry breaking is that the would-be Goldstone
bosons interact weakly at low energies and Effective Field Theory
Methods such as Chiral Perturbation Theory
(ChPT)~\cite{Gasser:1983yg,Gasser:1984gg} (for a review see
e.g. \cite{Pich:1995bw} and references therein), can successfully be
applied in terms of unknown low energy constants (LEC's) which cannot
be explained on the basis of the symmetry alone. This implies
neglecting the explicit effects of states about the mass gap which in
the case of two flavours might be identified with the scalar or vector
meson masses, and limits the maximum energy at which standard ChPT may
confidently be applied. Actually, the bulk of the values of the LEC's
can be saturated by the low energy contribution stemming from the
exchange of resonances located in the mass gap region. At finite, but
low, temperatures the physics of QCD is believed to consist of a gas
of heated hadrons and one still expects the dominant role to be played
by the pseudoscalar mesonic thermal excitations~\cite{Gasser:1986vb,
Gerber:1988tt,Smilga:1996cm}; effects of resonances are exponentially
suppressed by a Boltzmann factor in the mass gap $e^{-m_\rho/T}$.
Obviously, the applicability of such an approach requires that there
still be a mass gap and that confinement still holds. This entitles,
in particular, to consider the LEC's of the Chiral Lagrangian as
temperature independent couplings and finite (low) temperature model
independent predictions are
deduced~\cite{Pelaez:1998vx,Kaiser:1999mt,Pelaez:2002nd,Pelaez:2002xf,
GomezNicola:2004gg,GomezNicola:2002tn}.  From this point of view
finite temperature ChPT provides a strong theoretical constraint on
the QCD physics well below the phase transition. Extrapolations based
on ChPT suggest a melting of the condensate. However, it is unclear
whether this vanishing meets the very requirement of a phase
transition regarding the quark condensate.  Moreover, such a purely
hadronic based description can never account, by construction, for the
deconfinement phase transition expected from lattice
simulations~\cite{Karsch:1998hr}.

As already mentioned, the deconfinement phase transition can be
characterized for infinitely heavy quarks by a breaking of the center
symmetry of the colour gauge group. The interplay between this center
symmetry and chiral symmetry requires explicit consideration of quark
degrees of freedom and can quantitatively be assessed in chiral quark
models where the spontaneous chiral symmetry breaking is implemented
(see e.g. Ref.~\cite{RuizArriola:2002wr} for a review and references
therein) and the mass gap can be identified as twice the constituent
quark mass $M$. Chiral quark model Lagrangians are invariant under the
flavour chiral group $SU_R (N_f) \otimes SU_L ( N_f) $ and have often
been used to provide some semiquantitative understanding of hadronic
features in the low energy domain. In addition, at zero and finite
temperature the standard chiral quark models are invariant under {\em
global} $\text{SU}(N_c)$ transformations. At some critical temperature
Chiral quark models predict already at the one loop level a chiral
phase transition~\cite{Bernard:1987ir,Christov:1991se} at realistic
temperatures. Pion corrections where first considered in
Ref.~\cite{Florkowski:1996wf} (for a review see
e.g. Ref.~\cite{Oertel:2000jp}).  This has been traditionally
considered a big phenomenological success for these models, but they
suffer from the unphysical contribution of states which are not colour
singlets, so that even at lowest temperatures the hot environment
corresponds to a plasma of multiquarks, characterized by Boltzmann
factors $\sim e^{-n M /T} $ and not colour neutral hadronic
states~\cite{Megias:2004hj}. This unphysical feature can be avoided by
noting the relevance of large and local $\text{SU}(N_c)$ gauge
invariance at finite temperature. Basically, this corresponds to
include non-perturbative finite temperature gluons. Based on previous
works~\cite{Gocksch:1984yk,Meisinger:1995ih,Fukushima:2003fw} (see
also \cite{Ratti:2005jh,Ghosh:2006qh}) we have discussed at
length~\cite{Megias:2004hj} how large gauge invariance can efficiently
be implemented by the coupling of the Polyakov Loop to the standard
chiral quark models hence providing a cooling mechanism to the
chiral-deconfinement phase transition and the melting of the chiral
condensate is shifted up from $150 \,{\rm MeV}$ to about $250 \,{\rm MeV}$
in more quantitative agreement with QCD lattice simulations. The
coupling of QCD distinctive order parameters at finite temperature to
hadronic properties has been the subject of much attention over the
recent past \cite{Sannino:2002wb,Mocsy:2003qw,
Fukushima:2003fw,Fukushima:2003fm,Gocksch:1984yk,Meisinger:2003uf}.
Effective actions for the Polyakov loop as a confinement-deconfinement
order parameter have been proposed because of their relevance in
describing the phase transition from above the critical
temperature~\cite{Dumitru:2001xa,Meisinger:2001cq,
Dumitru:2003hp,Meisinger:2003id}.

In the present paper we want to analyze in more detail
these Polyakov-Chiral Quark Models and, more specifically, how the
Polyakov loop couples to the lightest Goldstone bosons at low
energies, and what is the net effect at low temperatures in the
effective Chiral Lagrangian.  Amazingly, the inclusion of Polyakov
loops, a colour source, at a quantum level reproduces the expectations
of low temperature ChPT based on the existence of a mass gap, namely
the tree level coefficients of the chiral Lagrangian are exponentially
suppressed in the mass gap. While this supports the ChPT-expected
dominance of pion fluctuations at low temperatures, the full model
unlike ChPT still predicts a rapid decrease of the mass gap at higher
temperatures as well as a sudden rise of the Polyakov loop at the {\it
same} temperature.

The paper is organized as follows. In Sect.~\ref{sec:ChQM_FT} we
describe shortly the Polyakov-Chiral quark models, and more
specifically the simplest version of them.  In Sect.~\ref{sec:chpt} we
set up the calculational framework of the low energy chiral Lagrangian
at finite temperature and display already the general structure of the
main result. In Sect.~\ref{sec:hk} we use the technique of heat kernel
expansion at finite temperature with Polyakov loops and the associated
large gauge invariance is preserved. This enables the calculation of
the effective Lagrangian in terms of the Polyakov loop and meson
pseudoscalar mesons as basic and independent degrees of freedom. In
Sect.~\ref{sec:integration} we go further and integrate over the
Polyakov loop variable in a gauge invariant manner, providing the form
of the effective (tree level) Lagrangian of ChPT at finite
temperature. Finally, in Sect.~\ref{sec:concl} we summarize our
conclusions. 

\section{Polyakov-Chiral quark models} 
\label{sec:ChQM_FT}

In this section we review the coupling of the Polyakov loop to chiral
quark models as a quantum and independent variable as suggested in
previous work \cite{Gocksch:1984yk,Meisinger:1995ih,%
Fukushima:2003fw,Megias:2004hj}. We follow \cite{Megias:2004hj} for
the basic ingredients involved in the construction of such
model. Since the coupling to quarks is rather universal we will
restrict ourselves to the simplest chiral model, namely, a constituent
quark model (CQ). The corresponding Lagrangian reads\footnote{In
Minkowski space we will use Bjorken-Drell conventions throughout the
paper.}
\begin{eqnarray}
{\cal L}_{\rm CQ} &=& \bar{q} (i\slashchar\partial 
+ \thru{v}^f + \thru{a}^f \gamma_5
-M U^{\gamma_5} - \hat{M}_0 )q \nonumber \\
&=:& \bar{q} \, i{\mathbf D} \, q \,, 
\label{eq:lagran_CQ}
\end{eqnarray} 
where $q=(u,d,s, \dots )$ represents a quark spinor with $N_c $ colors
and $N_f$ flavours. $\hat M_0= {\rm diag} (m_u, m_d, m_s,\ldots)$
stands for the current quark mass matrix. The symbols
$(v^f_\mu,a^f_\mu)$ denote external vector and axial-vector fields, in
flavour space. $M$ is the constituent quark mass and $U=e^{i \sqrt{2} \Phi /
f_\pi }$ ($f_\pi$ being the pion weak decay constant in the chiral
limit) is the flavour matrix representing the pseudoscalar octet of
mesons in the nonlinear representation:
\begin{eqnarray} 
	\Phi = \left( \matrix{ \frac{1}{\sqrt{2}} \pi^0 +
	\frac{1}{\sqrt{6}} \eta & \pi^+ & K^+  \cr  \pi^- & -
	\frac{1}{\sqrt{2}} \pi^0 + \frac{1}{\sqrt{6}} \eta & K^0  \cr 
	K^- & \bar{K}^0 & - \frac{2}{\sqrt{6}} \eta }
	\right) .
\end{eqnarray}
For vanishing current quark masses ${\cal L}_{\rm CQ}$ is invariant
under local ${\text U}(N_f)_R \otimes {\text U}(N_f)_L $
transformations. In addition there is a global SU($N_c$) symmetry.

After formally integrating out the quarks one gets the effective
action
\begin{equation}
\Gamma_{\rm CQ} = -i {\rm Tr} \log \left( i {\bf D} \right)\,.
\label{eq:eff_ac_njl} 
\end{equation} 
We use ${\rm Tr}$ for the full functional trace, $\tr_f $ stands for
the trace in flavour space, and $\tr_c $ for the trace in colour
space. The ultraviolate divergences introduced by the functional
determinant can be conveniently handled by the Pauli-Villars
method. Let us note that the issue of regularization is not so crucial
in the present case, $T\ll\Lambda$ \cite{Christov:1991se},
since the divergences affect only the zero temperature contributions
\cite{Kapusta:1989bk,LeBellac:1996bk}.

The chiral quark model coupled to the Polyakov loop corresponds to
introducing a non trivial colour component of the vector field in the
temporal direction
\begin{equation}
v_\mu^f \rightarrow v_\mu^f + v_\mu^c \,,
\quad
v_\mu^c= \delta_{\mu 0} v_0^c
\label{eq:2.4}
\end{equation}
in the Dirac operator, eq.~(\ref{eq:lagran_CQ}). The field $v^c_0(x)$
acts as a chemical potential in color space \cite{Fukushima:2003fw}.
Upon Wick rotation to pass to imaginary time (which we use to
introduce finite temperature) this field gives rise to the Polyakov
loop
\begin{equation}
\Omega(\vec{x},x_4) = {\mathcal T} \exp\left(
i\int_{x_4}^{x_4+\beta} dx_4^\prime \,v_4^c(\vec{x},x_4^\prime) 
\right)
\end{equation}
where ${\mathcal T}$ indicates the Euclidean temporal ordering and
$\beta=1/T$. The Polyakov loop plays an important role as an order
parameter for the deconfinement transition. In the present context it
appears quite naturally, since as shown in
\cite{Garcia-Recio:2000gt,Megias:2002vr} a generic effective action
with gauge fields at finite temperature will depend not only on the
standard zero temperature local gauge covariant operators ${\cal
O}_n(x)$ constructed with the covariant derivative, but also on the
Polyakov loop $\Omega(x)$ as a new finite temperature gauge covariant
operator. Specifically, at the one loop level the effective Lagrangian
takes the generic form \cite{Megias:2002vr}
\begin{equation}
{\cal L}(x)= \sum_n \tr[f_n(\Omega){\cal O}_n] \,.
\end{equation}

In the Polyakov-chiral quark model the partition function takes the
form
\begin{equation}
Z = \int DU D\Omega \, e^{i \Gamma_G[\Omega]} 
e^{i \Gamma_{\rm CQ}[U,\Omega]} \,,
\label{eq:Z_njl} 
\end{equation} 
where $DU$ is the Haar measure over the flavour group SU$(N_f)$
(actually, the product of local Haar measures) and $D \Omega $ is
essentially the Haar measure of the colour group SU$(N_c)$ (see
however Sect. \ref{sec:integration}), $\Gamma_G $ is the effective
gluon action whereas $\Gamma_{\rm CQ}$ stands for the quark effective
action in (\ref{eq:eff_ac_njl}). The model is motivated by the
underlying idea that all zero temperature gluon degrees of freedom
have been integrated out to yield the constituent quark mass, leaving
unintegrated the Polyakov loop as the only specifically finite
temperature gluonic degree of freedom. Ideally $\Gamma_G[\Omega]$
would be the result of such a partial integration in pure
gluodynamics. The Polyakov loop integration implements the gauge
invariant integration over the $v_0^c$ field \cite{Reinhardt:1996fs}.
Since gluons are vector fields, no colour axial-vector component is
introduced in (\ref{eq:2.4}).

As discussed in \cite{Megias:2004hj} the integration over the Polyakov
loop coupled to quarks suppresses unwanted coloured quark states and
mimics the confinement mechanism within the quark model. Note that the
Polyakov loop measure $ D\Omega \, e^{i \Gamma_G[\Omega]} $ preserves
center symmetry, i.e., invariance under 't Hooft transformations
\cite{'tHooft:1979uj} under which $\Omega\to z\Omega$, with
$z^{N_c}=1$. This symmetry is preserved by hadrons but not by
quarks. In the Polyakov-chiral quark model picture, as a valence quark
propagates at finite temperature, one can at any place insert a path
starting and ending at same point $x$ which winds a number $n$ of
times around the thermal cylinder, picking up a factor
$\langle(-\Omega)^n\rangle$ that goes with the propagator factor
$e^{-|n|M/T}$ (modulo we subleading polynomial correc\-tions). In the
presence of sea quarks there are further thermal windings, giving
roughly (being $L$ the number of sea quark loops)
\begin{equation}
\langle(-\Omega)^{n+n_1+\cdots+n_L}\rangle 
e^{-(|n|+|n_1|+\cdots+|n_L|)M/T} \,.
\end{equation}
The average over the gauge group is that for a pure Yang-Mills theory
and so preserves triality in the confined phase that we are
considering, i.e., $n+n_1+\cdots+n_L$ must be a multiple of $N_c$. For
$N_c\ge 3$, the leading thermal corrections comes from contributions
of the form $n=-n_1=1$, $n_2=\ldots=n_L=0$. This describes a thermal
$q\bar{q}$ pair, with a propagator factor $e^{-2M/T}$. Taking into
account quark interactions the leading thermal corrections will be of
the type $e^{-m_\pi/T}$, being the pion the lightest meson.

A remaining technical point deserves comment. As noted above the
presence of the Polyakov loop in the effective action is an
inescapable consequence of having a gauge theory at finite
temperature. So in addition to the standard colour Polyakov loop,
there will be also Polyakov loops associated to the vector and
axial-vector flavour external fields. Such {\em chiral} Polyakov loops
take the form
\begin{equation}
\Omega_{R,L}(x) = 
{\mathcal T} \exp \left( i \int_{x_4}^{x_4+\beta} dx_4^\prime \, 
(v^f_4(\vec{x},x_4^\prime) \pm  a^f_4(\vec{x},x_4^\prime))
\right) \,,
\end{equation}
and they appear automatically in any gauge invariant one loop
computation at finite temperature. Such operators (as well as the {\em
vector} flavour Polyakov loop) would introduce interesting new effects
such as {\em center symmetry in flavour space}, preserved by mesons but
not necessarily by baryons, depending on the number of flavours. Since
spin 1 mesons are relatively heavy we do not expect the flavour
Polyakov loops to be essential in the low temperature regime to be
study in this work, so it will be disregarded in what
follows. Nevertheless they could play a more active role near the
transition temperature.

\section{The structure of the low energy chiral Lagrangian at finite temperature} 
\label{sec:chpt} 


As already shown in previous works at zero temperature
\cite{Espriu:1989ff,RuizArriola:1991gc,Bijnens:1992uz,Megias:2004uj}
(see e.g. \cite{Megias:2005fj} for an updated list of references)
chiral quark models may provide a quantitative and microscopic
understanding of the structure of the low energy effective Lagrangian
of ChPT for the pseudoscalar mesons at {\it tree level}, namely
providing numerical values for the leading $N_c$ contributions to the
low energy constants (LEC's). In such a framework, external currents
are minimally coupled at the level of the more microscopic quark
degrees of freedom. On top of this, meson loops would provide
subleading $1/N_c$ contributions to the LEC's in addition to the
standard unitarity corrections of ChPT~\cite{RuizArriola:1991bq}.

In this section we extend the zero temperature results to the finite
temperature case and consider also the influence of the Polyakov loop
in the scheme described in the previous section. More explicitly, the
partition function in (\ref{eq:Z_njl}) can be written as
\begin{equation}
Z = \int DU D\Omega \, e^{i \Gamma_G[\Omega]} 
e^{i \int d^4x {\cal L}_\Omega(x)}
\label{eq:Z_njl1} 
\end{equation}
and then
\begin{equation}
Z = \int DU  \, e^{i \int d^4x {\cal L}^*(x)} \,.
\label{eq:Z_njl2} 
\end{equation}
The Lagrangian ${\cal L}_\Omega(x)$ is obtained after integration of
the quarks, and depends on the fields $U(x)$ and $\Omega(x)$. Setting
$\Omega=1$ one obtains the ChPT Lagrangian at finite temperature
corresponding to the CQ model without coupling to the Polyakov
loop. In the general case ${\cal L}_\Omega(x)$ provides also the
coupling of Goldstone mesons to the Polyakov loop field, and
interesting subject by itself. It is not clear, however, how such a
Lagrangian could be formulated in a model independent way since in QCD
the Polyakov loop is renormalized and in general will lie outside the
SU($N_c$) manifold.

The computation of ${\cal L}_\Omega(x)$ can be carried out by
following the methods developed in
Ref.~\cite{Garcia-Recio:2000gt,Megias:2002vr} and already applied to
QCD in \cite{Megias:2003ui}. The procedure is detailed in the next
Section. It is worth noticing that ${\cal L}_\Omega(x)$, either
directly, after setting $\Omega=1$, or averaged over the Polyakov
loop, represents a one quark-loop result and hence a {\em quenched}
approximation to the chiral Lagrangian.

Integration over the Polyakov loop field in (\ref{eq:Z_njl1}) gives
the (unquenched) finite temperature ChPT Lagrangian of the model,
denoted ${\cal L}^*(x)$, which is the main purpose of this work. In
ChPT at finite temperature it is usually assumed that the
corresponding low energy constants are {\it temperature
independent}~\cite{Gasser:1986vb} (see also
Refs.~\cite{GomezNicola:2002tn,Dobado:2002xf}). This is a quite
natural assumption because the applicability of ChPT is based on the
existence of a mass gap between the Goldstone bosons and the rest of
the hadronic spectrum. For non-strange mesons the gap is provided by
the $\rho$ meson mass $M_V$, so one expects the temperature dependence
of the LEC's to be of the order of $e^{-M_V /T }$. In a chiral quark
model, however, the pseudoscalar mesons are composite particles of
constituent quarks of a mass $M$, and the finite temperature also
influences their microscopic quark substructure. As a consequence the
LEC's become temperature dependent of the order $e^{-2M/T}$. Our
calculation below makes these remarks quantitative and also provides
an understanding on how the Polyakov loop cooling mechanism works in
favor of the ChPT expectations at finite temperature.

Although the calculation is straightforward, reducing effectively to
evaluation of Dirac and flavour traces, it is technically involved so
we discuss here the structure of the final result for the general
case. More details will be elaborated in the next sections.

The low energy effective Lagrangian written in the Gasser-Leutwyler
\cite{Gasser:1983yg} notation and in Minkowski space reads as follows
(we often use an asterisk as upperscript for finite temperature
quantities),
\begin{widetext}
\begin{eqnarray}
\cL^*{}^{(0)}(x) &=& B^*
\,, \label{eq:lag_0} \\ 
\cL^*{}^{(2)}(x) &=& \frac{f^*_\pi{}^2}{4}\tr_f\left( \D_\mu U^\dagger\D^\mu U 
+\chi^\dagger U +\chi U^\dagger \right) \,, 
\label{eq:lag_2} \\
\cL^*{}^{(4)}(x) &=& L^*_1\left(\tr_f(\D_\mu U^\dagger \D^\mu U)\right)^2
+L^*_2\left(\tr_f(\D_\mu U^\dagger \D_\nu U)\right)^2 \nonumber \\ 
&&+L^*_3\tr_f(\D_\mu U^\dagger \D^\mu U \D_\nu U^\dagger
\D^\nu U) 
+\overline{L}^*_3 \tr_f(\D_0 U^\dagger \D_0 U \D_\mu U^\dagger
\D^\mu U) \nonumber \\ 
&&+L^*_4\tr_f(\D_\mu U^\dagger \D^\mu
U)\tr_f(\chi^\dagger U + \chi U^\dagger) \nonumber \\
&&+L^*_5\tr_f(\D_\mu U^\dagger \D^\mu U(\chi^\dagger U +
U^\dagger\chi)) 
+\overline L^*_5\tr_f(\D_0 U^\dagger \D_0
U(\chi^\dagger U + U^\dagger\chi)) 
+\overline L_5^{\prime *}
\tr_f(\chi \D_0\D_0 U^\dagger + \chi^\dagger \D_0\D_0U) \nonumber \\
&&+L^*_6(\tr_f(\chi^\dagger U +\chi U^\dagger))^2
+L^*_7(\tr_f(\chi^\dagger U -\chi U^\dagger))^2 \nonumber \\
&&+\overline L^{\prime *} \tr_f(U^\dagger \D_0\D_0 U - U\D_0\D_0
U^\dagger) \tr_f(\chi^\dagger U - \chi U^\dagger) \nonumber \\
&&+L^*_8\tr_f(\chi^\dagger U\chi^\dagger U +\chi U^\dagger \chi
U^\dagger) \nonumber \\ 
&&-i L^*_9\tr_f(F^R_{\mu\nu}\D^\mu U^\dagger
\D^\nu U +F^L_{\mu\nu}\D^\mu U \D^\nu U^\dagger) \nonumber \\ 
&&-i\overline L^*_9 \tr_f\left(E_i^R(\D_0 U^\dagger \D^i U -\D^i U^\dagger \D_0
U)+E_i^L(\D_0 U \D^i U^\dagger -\D^i U \D_0 U^\dagger)\right) \nonumber \\
&&-i \overline L_9^{\prime *}\tr_f(\D_0E_i^R U^\dagger \D^i U +\D_0E_i^L U \D^i
U^\dagger) \nonumber \\ &&+L^*_{10}\tr_f(U^\dagger F_{\mu\nu}^L U
F^{\mu\nu}{}^R) 
+H^*_1\tr_f((F_{\mu\nu}^R)^2+(F_{\mu\nu}^L)^2) 
+\overline H^*_1 \tr_f((E_i^R)^2+(E_i^L)^2) 
+H^*_2\tr_f(\chi^\dagger \chi) \,.
\label{eq:lag_4}
\end{eqnarray}
\end{widetext} 
Here, $\tr_f$ means flavour trace and we have introduced the standard
chiral covariant derivatives and gauge field strength tensors,
\begin{eqnarray} 
\D_\mu U &=& D_\mu^L U-U D_\mu^R =
\partial_\mu U-i V_\mu^L U +iU  V_\mu^R, 
\nonumber \\ 
 F_{\mu\nu}^r &=& i[ D_\mu^r, D_\nu^r] = 
\partial_\mu V_\nu^r -\partial_\nu V_\mu^r
-i [ V_\mu^r , V_\nu^r ],
\label{eq:3.6}
\end{eqnarray} 
with $r=L, R$, and $V_\mu^{R,L} = v_\mu^f \pm a_\mu^f$. Finally, as at
zero temperature, we have normalized the explicit chiral breaking
field $\chi=2B_0^* \hat M_0$ so that $\cL^*{}^{(2)}$ takes a
standard form. This introduces a LEC $B_0^* = |\langle \bar q q
\rangle^*| / f^*_\pi{}^2 $, where $\langle \bar q q \rangle^* $ is the
quark condensate for one flavour at finite temperature.

The general form of the chiral low energy effective Lagrangian,
Eq.~(\ref{eq:lag_4}) requires some remarks. As we see there are terms
which can be written as the zero temperature Lagrangian but with
temperature dependent effective couplings. In addition, we also have
new terms, which due to the finite temperature generated by a heat
bath at rest necessarily break Lorentz invariance. This feature was
already pointed out in \cite{Megias:2002vr}. The remarkable, not yet
understood, feature is that there appear less terms breaking the
Lorentz symmetry than one might naively suggest (for instance a term
of the type $\tr_f( \D_0 U^\dagger\D_0 U)$ is missing). Thus, some
accidental symmetry may be at work and it would be interesting to find
it explicitly.

Although the form of the Lagrangian is quite gene\-ral, the particular
values of the low energy coefficients obviously depend on the specific
model. We will consider two particular chiral quark models: the
constituent quark model and the spectral quark model
\cite{RuizArriola:2003bs}. The results for the latter model will be
presented in appendix~\ref{sec:results_sqm}, and we concentrate on the
CQ model in the following sections.

\section{Heat kernel expansion at finite temperature in the presence
of Polyakov loops}
\label{sec:hk}

The calculation of the effective chiral Lagrangian of
Eqs.~(\ref{eq:lag_0})-(\ref{eq:lag_4}) can be divided into several
steps. First, we construct a Klein-Gordon operator out of the Dirac
operator and its adjoint for the non-anomalous part of the effective
action. After a generalized proper time re\-presentation we are
naturally lead to a heat kernel of the Klein-Gordon operator, for
which a heat kernel expansion is particularly suited. This
identification allows to directly apply the results of
\cite{Megias:2003ui,Megias:2002vr} and to work out the traces after
applying some relevant matrix identities which reduce the number of
independent operators. Finally, the equations of motion are also
considered to take into account that the pion fields are on the mass
shell. In the present section we assume a fixed value of the Polyakov
loop. This provides ${\cal L}_\Omega(x)$ of (\ref{eq:Z_njl1}). In
Sect. \ref{sec:integration} the Polyakov loop integration will be
addressed, to yield the unquenched chiral Lagrangian ${\cal L}^*(x)$
of (\ref{eq:Z_njl2}).

In this section and the next one we revert to Euclidean space,
$x_4=ix_0$, since it is much more convenient for calculations at
finite temperature in the imaginary time formalism. 
The partition function
is numerically unchanged, but takes the form
\begin{equation}
Z= \int
DU \, e^{-\int d^4 x \, \cL^*(x)} \quad\text{(Euclidean)} \,.
\end{equation}
Likewise, the low energy coefficients are also unchanged. E.g.,
\begin{eqnarray}
\cL^*{}^{(0)}(x) &=& -B^* \quad\text{(Euclidean)}
 \\
\cL^*{}^{(2)}(x) &=& \frac{f^*_\pi{}^2}{4}\tr_f\left( \D_\mu U^\dagger\D_\mu U 
-\chi^\dagger U -\chi U^\dagger \right) \,, 
\nonumber
\end{eqnarray}
with the same values of $B^*$, $f_\pi^*$, etc.

\subsection{The effective Klein-Gordon operator} 

The Dirac operator appearing in the fermion determinant behaves
covariantly under chiral transformations. This implies that, in
principle, one has to consider both vector and axial-vector
couplings. A great deal of simplification is achieved if the
conventions of Ref.~\cite{Salcedo:2000hx,Salcedo:2000hp} are
considered, where it is shown that it suffices to carry out the
calculation in the simpler case of a vector-like coupling and then
reconstruct the total chiral invariant result in a suitable way.
 
Let us hence consider the following Dirac operator with a vector-like
coupling,\footnote{Dirac gammas in Euclidean space are taken
Hermitian, and $\gamma_\mu\gamma_\nu=\delta_{\mu\nu}+\sigma_{\mu\nu}$.}
\begin{equation}
{\bf D} = \thruu{D}+h, \qquad h=m+z \,,
\end{equation}
where $h$ includes the pion field $m$, which we take ${\cal O}(p^0)$
in the chiral counting, and the chiral symmetry breaking mass term $z$
which we take ${\cal O}(p^2)$. We also introduce the following useful
notation
\begin{eqnarray}
m_{LR} &=& M U \,, \qquad m_{RL} = M U^\dagger \,, \\
z_{LR} &=& \frac{1}{2B_0^*}\chi \,, \qquad   z_{RL} 
= \frac{1}{2B_0^*}\chi^\dagger \,,
\end{eqnarray}
where $\chi$ has been introduced after (\ref{eq:3.6}).
(In the notation of
\cite{Salcedo:2000hx,Salcedo:2000hp}, the symbol $m$ is to be
interpreted as $m_{LR}$ or $m_{RL}$ depending on its position in the
formula, and similarly for the other vector-like symbols.)
It is convenient to introduce the adjoint Dirac operator
\begin{equation}
{\bf D}^\dagger =  -\thruu{D}+h \,.
\end{equation}
This definition allows us to separate the effective action into
$\gamma_5$-even and $\gamma_5$-odd components, corresponding to normal
and abnormal parity processes, since they correspond to the real and
imaginary parts of the Euclidean effective action, respectively (see
e.g. \cite{Salcedo:1996qy}).

We will focus on the normal parity component of the effective action,
which from Eq.~(\ref{eq:eff_ac_njl}) is formally given by
\begin{equation}  
\Gamma_{\rm CQ}^+ =-\frac{1}{2}\Tr\log({\bf D}^\dagger {\bf D})
=: \int_0^\beta dx_4
\int d^3{\mathbf x} \,{\cal L}_\Omega(x)\,,
\end{equation}
with the relevant Klein-Gordon operator
\begin{equation}
{\bf D}^\dagger {\bf D} =
-D^2_\mu+\frac{i}{2}\sigma_{\mu\nu}F_{\mu\nu}-\gamma_\mu{\D}_\mu h +
z^2 + \{m,z\}+M^2 \,, 
\end{equation}
We use the notation $\mathbf{D}_\mu h = [D_\mu , h]$. The field
strength tensor is defined as $F_{\mu\nu}=i[D_\mu, D_\nu]$. The
operator ${\bf D}^\dagger {\bf D}$ is of the Klein-Gordon type, with
mass term 
\begin{equation}\frac{i}{2}\sigma_{\mu\nu}F_{\mu\nu}-\gamma_\mu{\D}_\mu h
+ z^2 + \{m,z\}+M^2 \,.
\end{equation}
Therefore a heat kernel expansion becomes appropriate.  After a
(generali\-zed) proper-time regularization, the effective Lagrangian
in Euclidean space becomes
\begin{eqnarray}
{\cal L}_\Omega(x) &=& \frac{1}{2}\int_{0}^\infty
\frac{d\tau}{\tau}\phi\tau \,\tr \, 
\langle x|e^{-\tau {\bf D}^\dagger {\bf D}}|x \rangle 
\label{eq:lagran} \\
&=& \frac{1}{2}\int_{0}^\infty \frac{d\tau}{\tau}\phi(\tau)\frac{e^{-\tau
M^2}}{(4\pi\tau)^2}\sum_n (\tau)^n \tr(\hb_n(x)) \,.
 \nonumber 
\end{eqnarray}
This proper-time representation has been chosen to accommodate the
Pauli-Villars regularization as used in e.g.
Ref.~\cite{RuizArriola:1991gc}, 
\begin{eqnarray}
\phi(\tau) = \sum_i c_i e^{-\tau\Lambda_i^2} \,, 
\quad
c_0 &=& 1 \,, ~ \Lambda_0=0 \,,
\label{eq:pv}
\end{eqnarray}
but still using full advantage of the heat kernel. The conditions
$\sum_i c_i \Lambda_i^n=0$, for $n=0,2,4$, allow to render finite
the logarithmic, quadratic and quartic divergencies, respectively. To ${\cal O}
(p^4) $ terms we obtain the following contributions for the thermal
Seeley-DeWitt coefficients \cite{Megias:2002vr,Megias:2003ui} after
the Dirac trace has been implemented,
\begin{widetext} 
\begin{eqnarray}
\hb_0 &=& 4 \varphi_0(\Omega) \,,\nonumber \\ 
\hb_{1/2} &=& 0 \,, \nonumber \\ 
\hb_1 &=& 
-4\varphi_0(\Omega)\big(\{m,z\}+z^2\big)\,, \nonumber \\ 
\hb_{3/2} &=& 0  \,, \nonumber \\ 
\hb_2 &=& 
2\varphi_0(\Omega)\left((m_\mu)^2
+\{m_\mu,z_\mu\}+\{m,z\}^2
+\frac{1}{3}F_{\mu\nu}^2\right)
+\frac{2}{3}{\overline\varphi}_2(\Omega)E_i^2
+{\cal O}(p^6) \,, \nonumber \\ 
\hb_{5/2}&=&
{\cal O}(p^5) \,, \nonumber \\ 
\hb_3 &=&-\frac{2}{3}\varphi_0(\Omega)\left(
m_\mu\{m_\mu,\{m,z\}\}
+\{m,z\}(m_\mu)^2 
-i \{F_{\mu\nu},m_\mu
m_\nu\}
+im_\mu F_{\mu\nu} m_\nu 
+\frac{1}{2}(m_{\mu\nu})^2 \right)
+\frac{1}{3}{\overline\varphi}_2(\Omega)(m_{4\mu})^2 
+{\cal O}(p^5)\,, \nonumber \\
\hb_{7/2} &=& {\cal O}(p^5) \,, \nonumber \\ 
\hb_4 &=& \frac{1}{6}\varphi_0(\Omega)\bigg((m_\mu)^4 +m_\mu (m_\nu)^2 
m_\mu - (m_\mu m_\nu)^2\bigg) +{\cal O}(p^5)\,. 
\label{eq:Seeley_coef}
\end{eqnarray}
\end{widetext} 
In these formulas $E_i = F_{4i}$ is the ``electric field'' and a
notation of the type $X_{\mu\nu\alpha}$ has been used to mean
$\mathbf{D}_\mu \mathbf{D}_\nu \mathbf{D}_\alpha X =
[D_\mu,[D_\nu,[D_\alpha,X]]]$, e.g.
$m_{4\mu}=\mathbf{D}_4\mathbf{D}_\mu m$,
$F_{\alpha\mu\nu}=\mathbf{D}_\alpha F_{\mu\nu}$. The functions
$\varphi_0$ and ${\overline\varphi}_2$ are defined in appendix
\ref{sec:traces_integrals}.

Note that, in principle, since $D_\mu$ contains both flavour and
colour gauge fields, the Polyakov loop and the field strength tensor
$F_{\mu\nu}$ (and derivatives) in the heat kernel coefficients will
also contain flavour and colour contributions. As explained before in
the Polyakov loop we retain only its colour part. In addition, in the
strength tensor and derivatives we keep only the flavour part. This
neglects some gluonic corrections. This kind of QCD corrections have
been considered for instance in \cite{Bijnens:1992uz} for the
Nambu--Jona-Lasinio (NJL) model. We disregard them in the present
treatment since they are not specific of finite temperature.

\subsection{Effective Lagrangian}
\label{sec:eff_lagr}

The effective Lagrangian can be written as
\begin{equation}
\cL_\Omega(x) = \cL_\Omega^{(0)}(x) + \cL_\Omega^{(2)}(x) 
+ \cL_\Omega^{(4)}(x) + \cdots \,,
\end{equation}
where $\cL_\Omega^{(n)}$ is of ${\cal O}(p^n)$. Making use of the
expression (\ref{eq:lagran}), the flavour trace of the Seeley-DeWitt
coefficients obtained in (\ref{eq:csdw}) and after evaluating the
proper-time integrals using the Pauli-Villars regularization (see
appendix~\ref{sec:traces_integrals}), we get for the Lagrangian
$\cL_\Omega(x)$ an expression formally identical to (the Euclidean
version of) that in (\ref{eq:lag_0}-\ref{eq:lag_4}), except that the
coefficients depend on the Polyakov loop, that is, $B^*(\Omega)$
instead of $B^*$, etc.

At zeroth order one finds
\begin{equation}
B^*(\Omega) = -\frac{2N_f M^4}{(4\pi)^2} \tr_c \I_{-4}(\Omega) \,.
\label{eq:ini1}
\end{equation}
The second order gives
\begin{equation}
f^*_\pi{}^2(\Omega) = \frac{M^2}{4\pi^2} \tr_c\I_0(\Omega) \,.
\end{equation}
Note that we have not yet averaged over the Polyakov loop and the
(Polyakov loop independent) normalization constant $B_0^*$ (needed to
fix the definition of $\chi$) is determined after this average, as we
will explain in the next section. However the combination
$\chi/2B^*_0=\hat M_0$ is unambiguous. It contributes to
$\cL_\Omega^{(2)}$ (Euclidean) with
\begin{equation}
-\frac{M^3}{(4\pi)^2 B_0^*}\tr_c \I_{-2}(\Omega) 
\tr_f (\chi^\dagger U + U^\dagger \chi) 
\,,
\end{equation}
For the fourth order one obtains
\begin{eqnarray}
L^*_1 (\Omega) &=& \frac{1}{24(4\pi)^2}\tr_c\I_4(\Omega) \,, 
\label{eq:NJL_coef_L1o}\\
L^*_2(\Omega) &=& 2L^*_1(\Omega) \,, \\ 
L^*_3(\Omega) &=& -8L^*_1(\Omega) +\frac{1}{2}L^*_9(\Omega) \,, \\
\overline{L}_3^*(\Omega) &=&-\frac{1}{6(4\pi)^2}\tr_c\oI_2(\Omega) \,, \\ 
L^*_4(\Omega) &=& 0 \,, \\ 
L^*_5(\Omega) &=&
\frac{M}{2B^*_0}\left(\frac{f^*_\pi{}^2(\Omega)}{4M^2}-3L^*_9(\Omega)\right) \,, \\
\overline{L}_5^*(\Omega) &=& 
\overline{L}_5^{\prime *}(\Omega) = \frac{1}{2}\overline{L}_3^*(\Omega) \,, \\
L^*_6(\Omega) &=& 0 \,, \\ 
L^*_7(\Omega) &=& 
\frac{1}{8N_f} \left(-\frac{f^*_\pi{}^2(\Omega)}{2B_0^* M}+L^*_9(\Omega)\right)
\,, \\ 
\overline{L}^{\prime *}(\Omega) &=& 
-\frac{1}{4N_f}\overline{L}_3^*(\Omega) \,, \\
L^*_8(\Omega) &=& 
\frac{1}{16B_0^*}\left(\frac{1}{M}-\frac{1}{B_0^*}\right)f^*_\pi{}^2(\Omega)
-\frac{1}{8}L^*_9(\Omega) \,, \\ 
L^*_9(\Omega) &=&
\frac{1}{3(4\pi)^2}\tr_c\I_2(\Omega) \,, \\
\overline{L}_9^*(\Omega) &=& 
\overline{L}_9^{\prime *}(\Omega) = -\overline{L}_3^*(\Omega) \,, \\ 
L^*_{10}(\Omega) &=& -\frac{1}{2}L^*_9(\Omega) \,, \\ 
H_1^*(\Omega) &=&-\frac{f^*_\pi{}^2(\Omega)}{24M^2} +\frac{1}{4}L^*_9(\Omega) \,, \\ 
\overline{H}_1^*(\Omega) &=&-\frac{1}{6(4\pi)^2}\tr_c\oI_0(\Omega) \,, \\
H_2^*(\Omega) &=& -\frac{f^*_\pi{}^2(\Omega)}{8B^*_0{}^2}+\frac{1}{4}L^*_9(\Omega)\,.
\label{eq:fin1}
\end{eqnarray}
Note that all new Lorentz breaking terms, except $\overline{H}_1^*$,
are proportional among them. On the other hand, the standard
Gasser-Leutwyler coefficients can be expressed in terms of
$f^*_\pi{}^2, B_0^*, L^*_1$ and $L^*_9$ or, equi\-valently, in terms
of the integrals $\tr_c\I_n$ for  $n=-2,0,2,4$, which are computed in appendix
\ref{sec:traces_integrals}.

\section{Polyakov loop integration}
\label{sec:integration}

\subsection{Colour Group averaging} 

In order to proceed to the computation of the chiral (Euclidean) Lagrangian we
need to carry out the Polyakov loop integration,
\begin{equation}
e^{-\int d^4x {\cal L}^*(x)}= \int  D\Omega \, e^{- \Gamma_G[\Omega]} 
e^{- \int d^4x {\cal L}_\Omega(x)}
\label{eq:ZT}
\end{equation}
To this end we can take advantage of the chiral counting to expand the
exponential with ${\cal L}_\Omega(x)$ to fourth order. There is an
obstruction, however, since ${\cal L}_\Omega^{(0)}$ will appear to all
orders, making it difficult to take the Polyakov loop average. To sort
this problem we introduce a further {\em thermal} counting, in
addition to the chiral one. The thermal counting suppresses terms
which become irrelevant at low temperatures and so it is compatible
with the chiral expansion.

The thermal counting is defined as follows. ${\cal L}_\Omega(x)$ comes
from one quark-loop. As noted at the end of section \ref{sec:ChQM_FT},
any such loop will wind $\ell$ times around the thermal cylinder (in
imaginary time). Paths with $\ell=0$ give the zero temperature Lagrangian
${\cal L}_0$ which depends on $U$ but not on $\Omega$. The other paths
pick up a factor $\tr_c[(-\Omega)^\ell]$ which is accompanied by a
suppression factor (at low temperature) of the order of $e^{-|\ell|M/T}$
from the quark propagator. They give the thermal component of the
Lagrangian.
\begin{equation}
{\cal L}_\Omega(x)={\cal L}_0(x)+\sum_{\ell\ge 1}{\cal L}_\ell(x) \,.
\end{equation}
In this thermal counting the terms are ordered by the number $\ell$ of
Polyakov loop they carry with $\tr_c[(-\Omega)^\ell]$ as order
$|\ell|$. The discussion of the incidence of the gluonic part of the
action $\Gamma_G[\Omega]$ on this thermal expansion will be postponed
till Sec.~\ref{sec:gluonic_corrections}.

Combining the chiral and thermal expansions, one has
\begin{equation}
{\cal L}_\Omega(x)={\cal L}_0(x)
+\sum_{n,\ell}{\cal L}_\ell^{(n)}(x) \,.
\end{equation}
(Note that the chiral expansion of the zero temperature component
${\cal L}_0(x)$ is not required since it does not depend on the
Polyakov loop.) This is to be introduced in the Boltzmann factor
exponential of (\ref{eq:ZT}) and expanded. We will retain terms of
${\cal O}(p^4)$ in the chiral expansion and of leading order in the
thermal expansion.  Because the Haar measure $D\Omega$ (and actually
also the gluonic correction) preserves center symmetry, the first
thermal correction with a single quark loop (i.e., keeping just one
Lagrangian in the expansion of the exponential) will be of order
$\ell=N_c$ (a baryon-loop like contribution). For $N_c\ge 3$ this is
not the dominant contribution. The leading thermal correction comes
instead from correlation of a quark loop with an anti quark loop, a
meson-loop like configuration. Schematically, these corrections have thus the
following structure
\begin{equation}
{\cal L}_1^{(0)}{\cal L}_1^{(0)}+
{\cal L}_1^{(0)}{\cal L}_1^{(2)}+
\left( {\cal L}_1^{(0)}{\cal L}_1^{(4)}+
{\cal L}_1^{(2)}{\cal L}_1^{(2)}
\right)
\label{eq:5.4}
\end{equation}
contributing to $\cL^*{}^{(0)}$, $\cL^*{}^{(2)} $ and $\cL^*{}^{(4)}$,
respectively. Note that the two quark loops occur at different spatial
points, so the correlation between two Polyakov loops is needed. This we
model as \cite{Megias:2004hj}
\begin{eqnarray}
\int D\Omega \,\tr_c \Omega (\vec{x}) \, \tr_c
\Omega^{-1} (\vec{y}) = e^{-\sigma |\vec{x}-\vec{y}| /T} \,, 
\label{eq:corr-func}
\end{eqnarray} 
where $\sigma=(425\,\text{MeV})^2$ is the string tension. (This, of
course, implies that $D\Omega$ is not exactly the product of local
Haar measures.) This prescription is consistent with the zero
temperature quark-anti quark potential at medium and large
distances,\footnote{The short distance Coulombian part of the
$q\bar{q}$ potential cannot be reproduced with unrenormalized Polyakov
loops lying on the group manifold.} while at coincident points it
reproduces the SU($N_c$) group integration formula
\cite{Creutz:1984mg},
\begin{eqnarray}
\int d\Omega \, \tr_c \Omega \,  \tr_c \Omega^{-1} =1  \,.
\label{eq:7.5}
\end{eqnarray}
Effectively, the consequence of using (\ref{eq:corr-func}) in a
two-point computation of type (\ref{eq:5.4}) is to introduce a
correlation domain of spatial size $V$
\begin{eqnarray}
&&\int d^4 x \, d^4 y 
\int D\Omega \,\tr_c \Omega (\vec{x}) \, 
\tr_c \Omega^{-1} (\vec{y})
= \int d^4 x \frac{V}{T} \,,
\nonumber \\ &&
V= \frac{8 \pi T^3}{\sigma^3}  \,.
\label{eq:rule-vol}
\end{eqnarray}

Application of the procedure just described produces, for the zeroth
order of the effective Lagrangian in the chiral expansion,
\begin{equation}
B^* = B
+ \frac{N_f^2M^4T^3V}{\pi^4} K_2^2 \,.
\label{eq:L0_corT}
\end{equation}
In this formula $K_2=K_2(M/T)$ and $K_n(z)$ refers to the $n$-th order
Bessel function \cite{Abramowitz:1970bk}. We use the same convention
in all forthcoming expressions for the LEC. Not surprisingly, the
thermal correction to the vacuum energy density is proportional to the
correlation volume $V$ and contains exactly two Bessel functions from
the two correlated quark loops. This holds also for all LEC to follow.

To ${\cal O}(p^2)$ we find
\begin{eqnarray}
f^*_\pi{}^2 &=& f_\pi^2 -\frac{N_f M^4 T V}{\pi^4} 
 K_0 K_2 \,, \label{eq:fpi_corT}\\
B_0^* &=& B_0 + \frac{N_f M^4 T V}{\pi^4 f_\pi^2} K_2 
\left( B_0 K_0 - 2T K_1 \right) \,. 
\label{eq:B0_corT}   
\end{eqnarray}
And therefore, for the single flavor quark condensate at finite temperature
\begin{eqnarray}
\langle \bar{q}q \rangle^* &=& 
\langle \bar{q}q \rangle + \frac{2N_fM^4T^2V}{\pi^4}K_1 K_2
\end{eqnarray}

At very low temperatures, the asymptotic form of the Bessel
function~\cite{Abramowitz:1970bk}
\begin{eqnarray}
K_n (z) \sim  \left(\frac{\pi}{2z}\right)^{1/2}e^{-z}  \,,
\label{eq:bess_asy} 
\end{eqnarray} 
applies, and so the first thermal correction in the low energy
coefficients has the exponential suppression~$e^{-2M/T}$. This
indicates that meson thermal loops dominate at low temperatures (in
our treatment we ignore completely the quark binding effects so the
nature of the mesons involved cannot be resolved within the present
approach). Note that the group integration produces an $N_c$
suppression with respect to the zero temperature
contribution~\cite{Megias:2004hj}.\footnote{Naively, a two quark-loop
term would be ${\cal O}(N_c^2)$, and in fact, ignoring the Polyakov
loop dependence, i.e., setting $\Omega=1$ as in the standard NJL
model, would give $N_c^2$ , instead of unity on the r.h.s. of
(\ref{eq:7.5}). Thus the introduction of the Polyakov loop proves
essential to reproduce the ${\cal O}(N_c^0)$ result that one should
expect for the thermal corrections coming from colourless hadronic
excitations.}  We warn, however, that the results obtained with the
full Bessel functions and with their asymptotic form differ beyond
very low temperatures, especially as $n$ increases.

Due to center symmetry of the Polyakov loop measure, the higher
thermal contributions are of the general type $\int D\Omega \,
\tr_c(\Omega^{n_1})\cdots\tr_c(\Omega^{n_\ell})$, with $\sum_i n_i$ a
multiple of $N_c$. Among other, there are
\begin{itemize}
\item Two or more meson-like loops, of the type $\int D\Omega
\,(\tr_c \Omega \, \tr_c \Omega^{-1})^n$, $n=2,3,\ldots$, with a thermal
suppression factors $e^{- 2nM/T}$.
\item Baryon-like loop excitations, of the type $\int D\Omega \,\tr_c
\Omega^{\pm N_c}$, with thermal suppression~$e^{-N_c M/T}$. They come
from a single quark looping $N_c$ times the thermal cylinder. These
are the leading thermal corrections of hadronic type in the {\em
quenched} version of the theory \cite{Megias:2004kc,Megias:2005qf}.
\item Gluonic corrections, with thermal suppression~$e^{-m_G/T}$, $m_G
\sim 650 \, \text{MeV}$, from $\Gamma_G[\Omega]$.
\end{itemize}

In passing, we note that the leading thermal contribution to the
Polyakov loop expectation value can also be easily computed. It comes
from the coupling of the Polyakov loop observable with a quark
loop. Schematically, from $\langle\Omega\,\cL_1^{(0)}\rangle$. An easy
computation gives, at leading order
\begin{equation}
\langle \frac{1}{N_c}\tr_c\Omega\rangle = \frac{N_f}{N_c}\frac{M^2TV}{\pi^2}K_2 \,.
\end{equation}
This is a measure of the explicit breaking of center symmetry by
quarks, and is controlled by the ratio $M/T$ in the exponential
factor.

\subsection{Low Energy Coefficients}

Let us come now to ${\cal O}(p^4)$ terms. The result we find for the
Gasser-Leutwyler coefficients for the constituent quark model reads
\begin{eqnarray}
L_1^* &=& L_1 + \frac{M^4 V}{64\pi^4T}\left(
K_0^2-\frac{N_f}{6}K_2^2 \right) \,, \\ 
L_2^* &=& L_2 - 
\frac{N_fM^4 V}{192\pi^4T}K_2^2 \,, \\ 
L_3^* &=& L_3 + \frac{N_f M^3 V}{48\pi^4 T} K_2 
\left( M K_2 - T K_1 \right) \,,\\ 
\overline{L}_3^* &=& \frac{N_f M^4
V}{48\pi^4 T} K_0 K_2 \,, \\ 
L_4^* &=& \frac{M^4 V}{16\pi^4B_0} K_0 K_1 \,, \\ 
L_5^* &=& L_5-\epsilon L_5 +
\frac{N_f M^3 V}{16\pi^4 B_0} K_2 
\left( M K_1 -2T K_0 \right) \,, \\
\overline{L}_5^* &=& \overline{L}_5^{\prime *} =
\frac{1}{2}\overline{L}_3^* \,, \\ 
L_6^* &=& \frac{M^4 T V}{64 \pi^4 B_0^2} K_1^2 \,, \\ 
L_7^* &=& L_7
+\frac{f_\pi^2}{16N_fM B_0}\epsilon
\nonumber \\ &&\quad
+ \frac{M^3 V}{192 \pi^4} K_2
\left( \frac{12T}{B_0} K_0 - K_1 \right) \,, \\ 
\overline{L}^{\prime *} &=& -\frac{1}{4N_f} \overline{L}_3^* \,, \\ 
L_8^* &=& L_8
+\frac{f_\pi^2}{16B_0}\left(\frac{2}{B_0}-\frac{1}{M}\right) \epsilon
+ \frac{N_f M^3 V}{192\pi^4} K_2 
\nonumber \\ &&\times
\bigg(
K_1 + \frac{12M T}{B_0} \left( \frac{1}{B_0}-\frac{1}{M}\right)
K_0 \bigg) \,, \\ 
L_9^* &=& L_9 - \frac{N_f M^3 V}{24\pi^4} K_1 K_2 \,, \\ 
\overline{L}_9^* &=&
\overline{L}_9^{\prime *} = -\overline{L}_3^* \,, \\ 
L_{10}^* &=&
-\frac{1}{2} L_9^* \,, \\ 
H_1^* &=& H_1 + \frac{N_f M^2 V}{96\pi^4} K_2
\left( 4T K_0 - M K_1 \right) \,, \\ 
\overline{H}_1^* &=& \frac{N_f M^3 V}{24\pi^4} K_1 K_2 \,, \\ 
H_2^* &=& H_2 +\frac{f_\pi^2}{4 B_0^2}\epsilon
\nonumber \\ &&
\qquad
+
\frac{N_f M^3 V}{96\pi^4} K_2 
\left(
\frac{12 M T}{B_0^2} K_0 - K_1 \right) \,,
\end{eqnarray}
where we have defined the quantity
\begin{equation}
\epsilon = (B_0^*-B_0)/B_0 \,.
\end{equation}
As usual, to bring the fourth order Lagrangian to the standard form
(\ref{eq:lag_4}) we have used the equation of motion for the
nonlinear $U$ field. This is obtained by minimizing~$\cL^*{}^{(2)}(x)$
(and hence after the average over the Polyakov loop) and reads
\begin{equation}
u_{\mu\mu}u +u_\mu u_\mu -B^*_0[u,z]+\frac{B_0^*}{N_f}\tr_f([u,z])=0 \,.
\label{eq:ecm}
\end{equation}
(where the vector-like notation has been used, with $u_{LR} = U$ and
$u_{RL} = U^\dagger$). The last term arises from the constraint
$\det(U)=1$, since we are considering a $\textrm{SU}(N_f)$ flavour
group.

Explicit expressions for the zero temperature coefficients appear
in~\cite{Megias:2005fj}. Once again, we observe that the finite
temperature corrections are $N_c$-suppressed as compared to the zero
temperature values, as expected from hadronic excitations, and this
requires a proper Polyakov loop average. The terms with $N_f$, and also
$L_7^*$ and $\overline{L}^{\prime *}$, are those coming from ${\cal
L}_1^{(0)}{\cal L}_1^{(4)}$ while those without come from ${\cal
L}_1^{(2)}{\cal L}_1^{(2)}$.  It is worth noticing that some
structures which were absent in the quenched approximation
$\cL^{(4)}_\Omega(x)$, e.g. $L^*_4$ and $L^*_6$, are allowed in the
unquenched result, from ${\cal L}_1^{(2)}{\cal L}_1^{(2)}$.

\subsection{Volume rule}
\label{sec:volume_rule}

Clearly, to carry out the functional integration over the Polyakov
loop indicated in (\ref{eq:ZT}), besides requiring a careful
specification of the Polyakov loop action, would be an exceedingly
demanding task. However, we can make systematic relations of the type
(\ref{eq:rule-vol}) using a simple model. This will allow us to go
beyond the leading thermal corrections and to include gluonic
corrections from $\Gamma_G[\Omega]$. Specifically, we assume that the
space-time is decomposed into correlation domains of size $V/T$, in
such a way that two Polyakov loops are completely correlated if they
lie within the same domain and are completely decorrelated
otherwise. So for a partition function of the form
\begin{equation}
Z= \int D\Omega \, e^{-\int d^4 x \, \cL(x)} 
:= e^{-\int d^4 x \, \cL^\prime(x)}
\label{eq:5.31}
\end{equation}
we take
\begin{equation}
Z= \int\prod_n d\Omega_n \, e^{-\sum_n\Gamma_n}
\end{equation}
where $n$ labels the correlation domain, $d\Omega$ is the Haar
measure, and $\Gamma_n=(V/T)\cL(x_n)$, $x_n$ lying in the domain $n$.
This gives
\begin{eqnarray}
Z&=& \int\prod_n d\Omega_n\,
\left(1-\sum_n\Gamma_n+\frac{1}{2} 
\sum_{n,n^\prime}\Gamma_n\Gamma_{n^\prime}+\cdots\right)
\nonumber \\
&=&
 1
-\sum_n\langle\Gamma_n\rangle
+\frac{1}{2} 
\sum_n\langle\Gamma_n^2 \rangle
+\frac{1}{2}\sum_{n\not=n^\prime}\langle\Gamma_n\rangle
\langle\Gamma_{n^\prime}\rangle+\cdots
\nonumber \\
\end{eqnarray}
where $\langle~\rangle$ indicates average of $\Omega$ over SU($N_c$).
Obvious rearrangement gives, finally
\begin{equation}
\cL^\prime(x) =
\langle\cL(x)\rangle
-\frac{1}{2}\frac{V}{T}\left(\langle\cL^2(x)\rangle - \langle\cL(x)\rangle^2
\right)
+\cdots
\end{equation}
This is the standard cumulant expansion. Equivalently, $\cL^\prime(x)$
can be read off from
\begin{equation}
e^{-\frac{V}{T}\cL^\prime(x)} =
\left\langle
e^{-\frac{V}{T}\cL(x)}
\right\rangle_{\text{SU}(N_c)}
\label{eq:5.35}
\end{equation}
In concrete configurations the r.h.s. can be computed numerically
(being the integration manifold very well behaved.) Points to be noted
are: i) The effective action is an extensive quantity. It is given by
$\int d^4x\,\cL^\prime(x)$ and not $(V/T)\cL^\prime(x)$. ii) If
$\cL(x)$ contains a piece $\cL_0(x)$ which is Polyakov loop
independent, this term simply adds in $\cL^\prime(x)$. It does not
appear in the $V$-dependent contributions. And, iii) previous results
of this section, e.g. (\ref{eq:rule-vol}), are correctly reproduced.

\subsection{Gluonic Corrections}
\label{sec:gluonic_corrections}

Up to now in (\ref{eq:ZT}) we have neglected the factor
$e^{-\Gamma_G}$ in the Polyakov loop measure. Following
\cite{Fukushima:2003fw} we adopt a simple model inspired on the
lattice strong coupling expansion \cite{Polonyi:1982wz,Gross:1983pk}.
This model has one parameter which is fitted in
\cite{Fukushima:2003fw} to reproduce the transition temperature in
gluodynamics in the mean field approximation. (A more refined fit
involving more terms is discussed in \cite{Ratti:2005jh}.) Because our
calculation does not rely on the mean field approach, our
implementation of the model is not identical to that in those works
and so refitting would probably be required. Nevertheless, this will
not be needed here to make low temperature estimates since the gluonic
corrections are much suppressed. Specifically, since we want to have
extensivity, we introduce a local Lagrangian
\begin{equation}
\Gamma_G[\Omega]=\int d^4x\, \cL_g(x)
\end{equation}
which is modeled as
\begin{equation}
\cL_g(x)=-\frac{6}{a^3}Te^{-\sigma a /T}|\tr_c(\Omega)|^2
\label{eq:G_potential_sce}
\end{equation}
with an adjustable parameter $a^{-1}=272\,\text{MeV}$. Triality is
preserved, but, unlike the pure Haar measure, at higher temperatures the
action favors center of the group values for the Polyakov loop. (The
spontaneous breaking of center symmetry in gluodynamics indicates that
a Lagrangian with a different form would be needed in the high
temperature regime.) At low temperature we can see an exponential
thermal suppression controlled by a gluonic mass $m_G=\sigma
a=664\,\text{MeV}$. 

The gluonic corrections can be accounted for by including $\cL_g(x)$
in the Lagrangian of the volume rule (\ref{eq:5.35}). The leading
gluonic correction is ${\cal O}(e^{-m_G/T})$ and goes to the vacuum
energy density. Using (\ref{eq:7.5}) un easily finds $\delta_g B^*=
-\langle \cL_g(x) \rangle = 6 a^{-3}Te^{-m_G /T}$. At low temperature
this would be comparable with the meson-loop like terms, however,
recall that the constituent quark mass $M$ (actually $M^*$) gets
reduced as the temperature approaches the chiral transition, thereby
enhancing the role played by mesonic loops.

The gluonic corrections to $f^*_\pi{}^2$, $B_0^*$ and the other low
energy coefficients $L_i^*$ are of ${\cal O}(e^{-(2M+m_G)/T})$, as is
easily verified, thus they can be neglected as compared with the
hadronic-loop corrections.

Instead of modeling gluonic corrections through a local Lagrangian,
another approach suggested by the volume rule (\ref{eq:5.35}) is to
remove the Polyakov loop action from the Lagrangian and instead, to
replace the Haar measure average over the colour group by a weighted
average, $d\Omega \to d\Omega e^{-\Gamma_G[\Omega]}$. In principle
such a weight could be obtained through a proper sampling of the
Polyakov loop distribution on the group manifold within lattice
computations of gluodynamics. At present this is not feasible due to
severe numerical and conceptual problems related to renormalization
issues. An advantage of this approach would be that expectation values
of the type $\langle f(\Omega(x))\rangle$ (a single point) would be
exact when the quarks are switched off. However, it misses the
normalization of the measure, which is needed for its contribution to
the vacuum energy density. From (\ref{eq:5.35}) it is clear that such
weight is equivalent to a local Lagrangian $\cL_g(x)= (T/V)\Gamma_G$.

\section{Conclusions}
\label{sec:concl} 

The previous results clarify the question that chiral quark model
practitioners might legitimately ask, namely, whether the tree level
low energy coefficients in the chiral Lagrangian do genuinely depend
on tempera\-ture or not. According to our present calculation this
temperature dependence at very small temperatures is tiny. It is
exponentially suppressed by a scale which is {\it exactly the mass
gap}, and becomes inessential on temperature scales below the
deconfining transition (the region of tempera\-tures where the
Polyakov cooling mechanism acts) if the Polyakov loop is first
introduced and colour singlet projection is subsequently carried out
in a gauge invariant manner. In a quark model without Polyakov loop
the temperature dependence of the low energy constants behaves as
$L_i^* - L_i \sim N_c e^{-M/T}$, due to spurious contributions of
colour non singlet multi-quark states while the Polyakov cooling,
i.e., the explicit suppression of colourful excitations provides
instead the behaviour $L_i^* - L_i \sim e^{-2 M/T}$ as expected from
ChPT arguments. Remarkably, this cooling mechanism produces also the
correct $N_c$ counting. As shown in previous work~\cite{Megias:2004hj}
this effect at low temperatures manifests quantitatively in the chiral
symmetry restoration-center symmetry breaking phase transition.

To see how the agreement of Polyakov Chiral Quark Models to the
theoretical assumptions of ChPT at finite temperature well below the
phase transition materializes in practice we have calculated the
chiral effective Lagrangian at finite temperature. As a by product the
interaction between Polyakov loop and Goldstone bosons may be analyzed
in some detail.  The resulting chiral Lagrangian can be decomposed
into a zero temperature like looking piece with temperature dependent
low energy constants and a new Lorentz breaking contribution with
novel structures generated by the heat bath at rest. We remind that
these calculations aim at describing the tree level effective action
of ChPT at finite temperature. In any case, the corresponding
temperature induced effects on the low energy constants at this level
of approximation exhibit the Polyakov cooling. In other words, below
the phase transition any tempera\-ture dependence on the tree level
low energy constants may be neglected. This is precisely the starting
assumption of ChPT theory, a purely hadronic theory, and as we have
shown it arises naturally when the local and quantum nature of the
Polyakov loop is taken into account.

\begin{acknowledgments}
This work is supported in part
by funds provided by the Spanish DGI and FEDER founds with grant
no. FIS2005-00810, Junta de Andaluc\'{\i}a grant no. FM-225 and
EURIDICE grant number HPRN-CT-2003-00311.
\end{acknowledgments}

\appendix 

\section{Flavour traces and proper-time integrals}
\label{sec:traces_integrals}

In this appendix we compute the flavour traces of the Seeley-DeWitt
coefficients of Eq.~(\ref{eq:Seeley_coef}), and the proper time
integrals which appear in Eq.~(\ref{eq:lagran}), in order to obtain
the effective Lagrangian of sec.~\ref{sec:eff_lagr}. Euclidean
signature is used throughout.

\subsection{Flavour traces and useful identities}  

For $N_f=3$ flavours we have the $\textrm{SU}(3)$ identity
\begin{equation}
\tr(ABAB)=-2\tr(A^2 B^2)+\frac{1}{2}\tr(A^2)\tr(B^2)+(\tr(AB))^2
\end{equation}
where $A$ and $B$ are $3 \times 3$ Hermitian traceless matrices. From
here one gets
\begin{widetext} 
\begin{eqnarray}
\tr_f((m_{\mu}m_{\nu})^2)&=&
-2\tr_f((m_{\mu})^4)+\frac{1}{2}\Big(\tr_f((m_{\mu})^2)\Big)^2
+(\tr_f(m_{\mu}m_{\nu}))^2  \,,\label{eq:trf1}\\
\tr_f((m_{4}m_{\mu})^2) &=& -2\tr_f((m_{4})^2(m_{\mu})^2)
+\frac{1}{2}\tr_f((m_{4})^2)\tr_f((m_{\mu})^2)+(\tr_f(m_{4}m_{\mu}))^2 \,.
\label{eq:trf2}
\end{eqnarray}
Further useful identities are  
\begin{eqnarray}
\tr_f((m_{\mu\nu})^2) &=& 
\tr_f((m_{\mu\mu})^2)
+2i\tr_f(F_{\mu\nu}m_\mu m_\nu)
-\tr_f((mF_{\mu\nu})^2)
+M^2\tr_f(F_{\mu\nu}^2) \,, 
\label{eq:trf3}\\
\tr_f((m_{4\mu})^2) &=& 
\tr_f(m_{44}m_{\mu\mu})
+2i\tr_f(E_i [m_4,m_i])
+2i\tr_f(E_{4i}m m_i) \,,
\label{eq:trf4}
\end{eqnarray}
where we have applied the identity $X_{\mu\nu} = X_{\nu\mu}
-i[F_{\mu\nu},X]$, being $X$ any operator. Using the equation of
motion, Eq.~(\ref{eq:ecm}), we obtain the following identities
\begin{eqnarray}
\tr_f(m_\mu z_\mu) &=&
\frac{1}{2B_0^*M^2}\tr_f((m_\mu)^2 mx)
-\frac{1}{4B_0^*M}\tr_f((mx)^2)
+\frac{M}{4B_0^*}\tr_f(x^2)
+\frac{1}{8MN_fB_0^*}(\tr_f([m,x]))^2
\,,
\label{eq:trf5} \\
\tr_f((m_{\mu\mu})^2) &=&
\frac{1}{M^2}\tr_f((m_\mu)^4)
-\frac{1}{2} \tr_f((m x)^2) 
+\frac{M^2}{2}\tr_f(x^2)
+\frac{1}{4N_f}(\tr_f([m,x]))^2 \,, 
\label{eq:trf6}\\
\tr_f(m_{44}m_{\mu\mu}) &=&
\frac{1}{M^2}\tr_f((m_4)^2 (m_\mu)^2)
-M\tr_f(m_{44}x) 
-\frac{1}{M}\tr_f((m_4)^2 mx)
+\frac{1}{2MN_f}\tr_f(m_{44}m)\tr_f([m,x]) \,,
\label{eq:trf7}
\end{eqnarray}
\end{widetext}
where the normalized field $x =2B_0^* z$ has been introduced and so
\begin{equation}
x_{LR} =\chi \,, \qquad x_{RL}=\chi^\dagger \,.
\end{equation}
Applying Eqs.~(\ref{eq:trf1})-(\ref{eq:trf7}) the flavour trace of the
Seeley-DeWitt can be worked out, yielding
\begin{widetext} 
\begin{eqnarray}
\tr_f \hb_0 &=& 4N_f \varphi_0(\Omega) \,, 
\nonumber \\ 
\tr_f\hb_1 &=& 
-\varphi_0(\Omega)\left(\frac{4}{B_0^*}\tr_f(mx)
+\frac{1}{B^*_0{}^2}\tr_f(x^2)\right) \,,
 \nonumber \\ 
\tr_f \hb_2 &=& 2\varphi_0(\Omega)\tr_f((m_\mu)^2) 
+\frac{2}{B_0^* M^2}\varphi_0(\Omega)\tr_f((m_\mu)^2 mx)
+\frac{1}{B_0^*}\left(\frac{1}{B_0^*}
-\frac{1}{M}\right)\varphi_0(\Omega)\tr_f((mx)^2) 
\nonumber \\
&& 
+\frac{M}{B_0^*}\left(\frac{M}{B_0^*}+1\right)\varphi_0(\Omega)\tr_f(x^2)
+\frac{2}{3}\varphi_0(\Omega)\tr_f(F_{\mu\nu}^2)
+\frac{2}{3}\overline\varphi_2(\Omega)\tr_f(E_i^2) +\frac{1}{2MN_f
B_0^*}\varphi_0(\Omega)(\tr_f([m,x]))^2 \,, \nonumber \\
\tr_f \hb_3 &=& 
\frac{4}{3}i\varphi_0(\Omega)\tr_f(F_{\mu\nu}m_\mu m_\nu)
+\frac{1}{3}\varphi_0(\Omega)\tr_f((mF_{\mu\nu})^2)
-\frac{1}{3}M^2\varphi_0(\Omega)\tr_f(F_{\mu\nu}^2)
-\frac{1}{6}M^2\varphi_0(\Omega) \tr_f(x^2) \nonumber \\
&&
+\frac{1}{6}\varphi_0(\Omega)\tr_f((mx)^2)
-\frac{2}{B_0^*}\varphi_0(\Omega)\tr_f((m_\mu)^2 mx)
-\frac{1}{3M}\overline\varphi_2(\Omega)\tr_f((m_4)^2mx)
-\frac{M}{3}\overline\varphi_2(\Omega)\tr_f(m_{44}x) \nonumber \\
&&
+\frac{2}{3}i\overline\varphi_2(\Omega)\tr_f(E_i[m_4,m_i])
+\frac{2}{3}i\overline\varphi_2(\Omega)\tr_f(E_{4i}mm_i)
-\frac{1}{3M^2}\varphi_0(\Omega)\tr_f((m_{\mu})^4)
+\frac{1}{3M^2}\overline\varphi_2(\Omega)\tr_f((m_4)^2(m_{\mu})^2)
\nonumber \\
&&
-\frac{1}{12N_f}\varphi_0(\Omega)(\tr_f([m,x]))^2
+\frac{1}{6MN_f}\overline\varphi_2(\Omega)\tr_f(m_{44}m)\tr_f([m,x])
) \,, \nonumber \\ 
\tr_f \hb_4 &=&
-\frac{1}{12}\varphi_0(\Omega)\Big(\tr_f((m_\mu)^2)\Big)^2
-\frac{1}{6}\varphi_0(\Omega)(\tr_f(m_\mu m_\nu))^2
+\frac{2}{3}\varphi_0(\Omega)\tr_f((m_\mu)^4) \,.
\label{eq:csdw}
\end{eqnarray}
\end{widetext}
 
\subsection{Proper-time integrals with Polyakov loop}

The basic proper-time integrals are defined by
\begin{eqnarray}
\I_{2l}(\Lambda,M,\Omega) &:=& M^{2l} \int_0^\infty \frac{d\tau}{\tau}
\phi(\tau)\tau^l e^{-\tau M^2} \varphi_0(\Omega) \,, \label{eq:I2l}
\\
\oI_{2l}(\Lambda,M,\Omega) &:=& M^{2l}\int_0^\infty \frac{d\tau}{\tau}
\phi(\tau)\tau^l e^{-\tau M^2} \overline\varphi_2(\Omega) \,,
\label{eq:oI2l}
\end{eqnarray}
where $\Omega$ is a $\textrm{SU}(N_c)$ matrix in colour space, and we
define $\varphi_0$ and $\overline\varphi_2$ as follows
\begin{eqnarray}
\varphi_0(\Omega) &=& \sum_{n\in\mathbb Z} 
e^{-\frac{n^2\beta^2}{4\tau}}(-\Omega)^n \,, \\
\overline\varphi_2(\Omega) &=&
\frac{\beta^2}{2\tau}\sum_{n\in\mathbb{Z}} n^2
e^{-\frac{n^2\beta^2}{4\tau}}(-\Omega)^n \,.
\end{eqnarray}
The $n=0$ term in the sum corresponds to the zero temperature
contribution, and for such a term the Pauli-Villars regularization can
be applied. In the remaining $n\ne 0$ terms the regularization will be
removed, since these thermal terms are ultraviolet finite. This
approximation is justified for temperatures well below the cutoff. The
calculation of the integral yields
\begin{widetext} 
\begin{eqnarray}
M^4\I_{-4}(\Lambda,M,\Omega) &=& 
-\frac{1}{2}\sum_i c_i (\Lambda_i^2+M^2)^2\log(\Lambda_i^2+M^2)
+8(MT)^2\sum_{n=1}^\infty\frac{K_2(n M/T)}{n^2} 
\big((-\Omega)^n+(-\Omega)^{-n}\big) \,, 
\nonumber \\ 
M^2 \I_{-2}(\Lambda,M,\Omega) &=& \sum_i c_i
(\Lambda_i^2+M^2)\log(\Lambda_i^2+M^2)
+4MT\sum_{n=1}^\infty\ \frac{K_1(n M/T)}{n}
\big((-\Omega)^n+(-\Omega)^{-n}\big) \,,
\nonumber \\ 
\I_0(\Lambda,M,\Omega)
&=& -\sum_i c_i \log(\Lambda_i^2+M^2) +
2\sum_{n=1}^\infty K_0(n M/T) \big((-\Omega)^n+(-\Omega)^{-n}\big) \,, 
\nonumber \\
\I_{2l}(\Lambda,M,\Omega) &=&\Gamma(l)\sum_i
c_i\left(\frac{M^2}{\Lambda_i^2+M^2}\right)^l 
+ 2\left(\frac{M}{2T}\right)^l\sum_{n=1}^\infty
n^l K_l(n M/T) \big((-\Omega)^n+(-\Omega)^{-n}\big) \,,
\qquad  \textrm{Re}(l)>0 \,, 
\nonumber \\ 
\oI_{2l}(\Lambda,M,\Omega) &=&
4\left(\frac{M}{2T}\right)^{l+1}\sum_{n=1}^\infty
n^{l+1} K_{l-1}(n M/T) \big((-\Omega)^n+(-\Omega)^{-n}\big) \,, 
\qquad l\in \mathbb{R} \,.
\label{eq:prop_time_int}
\end{eqnarray}
\end{widetext} 

\section{Results for the Spectral Quark Model}
\label{sec:results_sqm}

In this appendix we present the results of the low energy coefficients
for the Spectral Quark Model (SQM). In the SQM the effective action reads
\begin{eqnarray}
\Gamma_{\rm SQM} =-i\int d \omega \rho(\omega) {\rm Tr} \log
\left( i {\bf D} \right),
\label{eq:eff_ac_sqm} 
\end{eqnarray} 
where the Dirac operator is given by 
\begin{eqnarray}
i {\bf D} &=& i\slashchar{\partial} +\thru{v}+\thru{a}\gamma_5 
- \omega U^{\gamma_5} - {\hat M_0} 
\label{eq:dirac_op_sqm} 
\end{eqnarray} 
and $\rho(\omega)$ is the spectral function of a generalized Lehmann
representation of the quark propagator with $\omega$ the spectral mass
defined on a suitable contour of the complex plane
\cite{RuizArriola:2001rr,RuizArriola:2003bs,RuizArriola:2003wi,Megias:2004uj}.
The use of certain spectral conditions guarantees finiteness of the
action.

A judicious choice of the spectral function based on vector meson
dominance generates a quark propagator with no-poles (analytic
confinement).  More details of the SQM for at zero and finite
temperature relevant for the present paper are further developed at
Ref.~\cite{Megias:2004hj}. The partition function for the SQM can be
written as
\begin{eqnarray}
Z = \int DU D\Omega \, e^{i \Gamma_G[\Omega]} 
e^{i \Gamma_{\rm SQM}[U,\Omega]} \,.
\label{eq:Z_sqm}
\end{eqnarray} 

In this model one has to compute an
average over the constituent quark mass with an spectral weight
function $\rho(\omega) $ satisfying a set of spectral conditions 
Note that $M$
appears as argument in the integrals $\I_{2l}$ and $\oI_{2l}$, but
also as additional multiplicative factors in the Gasser-Leutwyler
coefficients. This generates a larger number of independent functions
as compared to the NJL model. 

We write first the results for the effective Lagrangian before the
group average. We get for the zero order Lagrangian
\begin{equation}
\cL^{(0)}_\Omega = -\frac{2N_f}{(4\pi)^2}\langle \omega^4
\tr_c\I_{-4}(\Omega)\rangle\,. \label{eq:ini2}
\end{equation}
To simplify the notation we indicate with $\langle \quad \rangle$ the
spectral average $\int d\omega \rho(\omega)$. The second order
Lagrangian becomes
\begin{eqnarray}
\cL^{(2)}_\Omega &=&
\frac{f^*_\pi{}^2(\Omega)}{4}\tr_f((u_\mu)^2) \nonumber \\
&& + \frac{2}{(4\pi)^2
B_0^*}\langle \omega^3\tr_c \I_{-2}(\Omega) \rangle \tr_f (ux) \,,
\end{eqnarray} 
with
\begin{equation}
f^*_\pi{}^2(\Omega) = \frac{1}{4\pi^2}\langle
\omega^2\tr_c\I_0(\Omega)\rangle \,. 
\end{equation}
As in the CQ model, $B_0^*$ will be obtained after the average in
$\Omega$. The fourth order Lagrangian has the low energy coefficients
\begin{eqnarray} 
L^*_1(\Omega) &=& \frac{1}{24(4\pi)^2}\langle \tr_c\I_4(\Omega)\rangle
\,, 
\nonumber \\ 
\overline{L}_3^*(\Omega) &=& -\frac{1}{6(4\pi)^2}\langle
\tr_c\oI_2(\Omega)\rangle \,, \nonumber \\ L^*_5(\Omega) &=&
\frac{1}{2(4\pi)^2B_0^*}\bigg(\langle\omega\tr_c\I_0(\Omega)\rangle
-\langle\omega\tr_c\I_2(\Omega)\rangle \bigg)\,, \nonumber \\ L^*_7(\Omega) &=&
\frac{1}{2(4\pi)^2N_f}\left(-\frac{1}{2B_0^*}
\langle\omega\tr_c\I_0(\Omega)\rangle+4\pi^2L^*_9(\Omega)\right)
\,, \nonumber \\ 
L^*_8(\Omega) &=&
\frac{1}{4(4\pi)^2B_0^*}\langle\omega\tr_c\I_0(\Omega)\rangle
-\frac{f^*_\pi{}^2(\Omega)}{16B^*_0{}^2} -\frac{1}{8}L^*_9(\Omega) \,,
\nonumber \\ L^*_9(\Omega) &=& \frac{1}{3(4\pi)^2}\langle
\tr_c\I_2(\Omega)\rangle \,, \nonumber \\ H_1^*(\Omega) &=&
-\frac{1}{6(4\pi)^2}\langle\tr_c\I_0(\Omega)\rangle +
\frac{1}{4}L^*_9(\Omega) \,, \nonumber \\ \overline{H}_1^*(\Omega) &=&
-\frac{1}{6(4\pi)^2}\langle\tr_c\oI_0(\Omega)\rangle \,, \nonumber \\
H_2^*(\Omega) &=& \frac{1}{2(4\pi)^2B_0^*}\left(\frac{1}{B_0^*}\langle
\omega^2\tr_c \I_{-2}(\Omega)\rangle-\langle
\omega\tr_c\I_0(\Omega)\rangle\right) \nonumber \\
&-&\frac{f^*_\pi{}^2(\Omega)}{8B^*_0{}^2}+\frac{1}{4}L^*_9(\Omega) \,.
\label{eq:fin2}
\end{eqnarray}
The remaining coefficients satisfy the same geometric relations
obtained in the NJL, Eqs.~(\ref{eq:NJL_coef_L1o})-(\ref{eq:fin1}). We
also get the relation
\begin{equation}
L^*_7(\Omega) =
-\frac{1}{N_f}\left(\frac{f^*_\pi{}^2(\Omega)}{16B^*_0{}^2}+L^*_8(\Omega)\right) \,,
\end{equation}
both in the SQM and in NJL. 

The next step is to carry out the integration over Polyakov loops, as
in Sec.~\ref{sec:integration}. For the temperature independent part
the result can be expressed in terms of the spectral log-moments as
was already done in Ref.~\cite{Megias:2004uj}. For the temperature
dependent contributions we will explicitly use the meson dominant form
of the spectral function $\rho(\omega)$. After computing all
avera\-ges (spectral and colour group average), we get for the zeroth
order effective Lagrangian at leading order in the thermal expansion
\begin{equation}
B^* = B + \frac{N_f^2 T^3 V}{\pi^4} \langle \omega^2 K_2\rangle^2 \,.
\end{equation}
In this expression $K_2=K_2(\omega/T)$. We use the same convention for
the symbol $K_n$ in subsequent expressions. The spectral average can be computed
and gives
\begin{equation}
\langle \omega^2 K_2\rangle 
= \frac{T^2}{24}(48+24 x_V + 6 x_V^2 + x_V^3) e^{-x_V/2} \,.
\end{equation}
We use the notation $x_V=M_V/T$. 

The pion weak decay constant and the normalization constant read
respectively
\begin{eqnarray}
\frac{f_\pi^*{}^2}{f_\pi^2} &=& 1 - \frac{N_f T V}{\pi^4 f_\pi^2} 
\langle \omega^2 K_0\rangle \langle \omega^2 K_2\rangle \,, \\
\frac{B_0^*}{B_0} &=& 1 + \frac{N_f T V}{\pi^4 f_\pi^2} 
\langle \omega^2 K_2\rangle 
\nonumber \\
&& \times \left(
\langle \omega^2 K_0\rangle - \frac{2 T}{B_0}
\langle \omega^2 K_1\rangle \right) \,,
\end{eqnarray}
with the spectral averages
\begin{eqnarray}
\langle \omega^2 K_0\rangle &=& \frac{T^2}{24} x_V^2(2+x_V) e^{-x_V/2} \\
\langle \omega^2 K_1\rangle &=& \frac{\rho_3^\prime}{4T} e^{-x_S/2} \,,
\end{eqnarray}
and $x_S = M_S/T$. The equations of motion are those in
Eq.~(\ref{eq:ecm}). (Note that $B_0^*$, and hence the normalization of
the field $\chi$ depends on the model.) The low energy coefficients,
including the first thermal correction, read
\begin{widetext}
\begin{eqnarray}
L_1^* &=& L_1  + 
\frac{2 V}{3(4\pi)^4 T}\bigg(6\langle \omega^2 K_0\rangle^2 -N_f 
\langle \omega^2 K_2\rangle^2\bigg) \,, \\
L_2^* &=& L_2  -\frac{N_f V}{3(8\pi^2)^2 T} 
\langle \omega^2 K_2\rangle^2 \,, \\
L_3^* &=& L_3 + \frac{N_f V}{3(2\pi)^4 T} 
\langle \omega^2 K_2\rangle \bigg( 
\langle \omega^2 K_2\rangle -T^2 \langle \omega K_1\rangle
\bigg) \,, \\
\overline{L}_3^* &=& \frac{N_f V}{3(2\pi)^4T} \langle \omega^2 K_0\rangle 
\langle \omega^2 K_2\rangle \,, \\
L_4^* &=& \frac{V}{(2\pi)^4 B_0} \langle \omega^2 K_0\rangle 
\langle \omega^2 K_1\rangle \,, \\
L_5^* &=& (1-\epsilon) L_5 - \frac{N_f V}{(2\pi)^4 B_0}
\langle \omega^2 K_2 \rangle 
\bigg(
2T \langle \omega K_0\rangle  - \langle \omega K_1\rangle
\bigg)\,, \\
\overline{L}_5^* &=& \overline{L}_5^{\prime *} = \frac{1}{2} \overline{L}_3^* \,, \\
L_6^* &=& \frac{TV}{64\pi^4 B_0^2} \langle \omega^2 K_1\rangle^2 \,, \\
L_7^* &=& L_7 + \frac{L_5}{2N_f}\epsilon
+\frac{V}{192\pi^4} \langle \omega^2 K_2\rangle 
\bigg(
\frac{12T}{B_0} \langle \omega K_0\rangle 
- \langle \omega^2 K_1\rangle
\bigg) \,, \\
\overline{L}^{\prime *} &=& -\frac{1}{4N_f}\overline{L}_3^* \,, \\
L_8^* &=& L_8 + \left(\frac{f_\pi^2}{8B_0^2}-\frac{L_5}{2}\right)\epsilon
+\frac{N_f V}{192 \pi^4} \langle \omega^2 K_2\rangle
\bigg(
\langle \omega^2 K_1\rangle 
+\frac{12T}{B_0}\left(\frac{1}{B_0}\langle \omega^2 K_0\rangle 
- \langle \omega K_0\rangle \right)
\bigg) \,, \\
L_9^* &=& L_9 -\frac{N_f T V}{24\pi^4} \langle \omega K_1
\rangle \langle \omega^2 K_2\rangle \,, \\
\overline{L}_9^* &=& \overline{L}_9^{\prime *} = - \overline{L}_3^* \,, \\
L_{10}^* &=& -\frac{1}{2} L_9^* \,, \\
H_1^* &=&  H_1 + \frac{N_f V}{6 (2\pi)^4} \langle \omega^2 K_2\rangle 
\bigg( 4T \langle K_0\rangle - \langle \omega^2 K_1\rangle
\bigg) \,, \\
\overline{H}_1^* &=& \frac{N_f V}{24\pi^4} 
\langle \omega K_1\rangle \langle \omega^2 K_2\rangle \,, \\
H_2^* &=& H_2 +\left(L_5+\frac{f_\pi^2}{2B_0^2}\right)\epsilon 
+\frac{N_f V}{96 \pi^4} \langle \omega^2 K_2\rangle 
\bigg[
\frac{12T}{B_0}\left(\langle \omega K_0\rangle 
+ \frac{1}{B_0}\langle \omega^2 K_0\rangle\right) \nonumber \\
&&- \langle \omega^2 K_1\rangle - \frac{24 T^2}{B_0^2} 
\langle \omega K_1\rangle
\bigg] \,,
\end{eqnarray}
\end{widetext} 
where $\epsilon = (B_0^*-B_0)/B_0$. The spectral averages are
\begin{eqnarray}
\langle K_0\rangle &=& 
-\frac{1}{2}\gamma_E - \log(x_V/4)+\frac{1}{2}\psi(5/2) 
\nonumber \\ &&
+\frac{x_V^5}{7200} \, {}_1 F_2\left[5/2;7/2,7/2;(x_V/4)^2\right]
\nonumber \\ &&
+\frac{x_V^2}{48} \,
{}_2 F_3\left[1,1;-1/2,2,2;(x_V/4)^2\right] \,, \\
\langle \omega K_0\rangle &=& 
\frac{\rho_3^\prime}{2T^2 x_S^2}(2+x_S) e^{-x_S/2} \,, \\
\langle \omega K_1\rangle &=& \frac{T}{12}(12+6x_V+x_V^2) e^{-x_V/2}
\end{eqnarray}
${}_p F_q [a_1,\dots,a_p;b_1,\dots,b_q;z]$ is the generalized
hypergeometric function. The expressions for the zero temperature
coefficients appear in \cite{Megias:2004uj}.


\end{document}